\documentclass[sigconf]{acmart}
\usepackage{multirow}
\usepackage{subcaption}
\usepackage{enumitem}
\usepackage{booktabs}

\usepackage{acmart-taps}

\AtBeginDocument{%
  }

\usepackage{amsmath,amssymb}
\usepackage{mathtools}
\usepackage{algorithm}
\usepackage{algpseudocode}

\usepackage{array}
\usepackage{multirow}
\usepackage{graphicx}
\usepackage{enumitem}
\usepackage{booktabs}
\usepackage{etoolbox}
\usepackage{dsfont}
\usepackage{balance}

\usepackage{booktabs}

\copyrightyear{2024} 
\acmYear{2024} 
\setcopyright{acmlicensed}\acmConference[RecSys '24]{18th ACM Conference on Recommender Systems}{October 14--18, 2024}{Bari, Italy}
\acmBooktitle{18th ACM Conference on Recommender Systems (RecSys '24), October 14--18, 2024, Bari, Italy}
\acmDOI{10.1145/3640457.3688140}
\acmISBN{979-8-4007-0505-2/24/10}

\acmPrice{15.00}

\begin{document}

\title{Scalable Cross-Entropy Loss for Sequential Recommendations with Large Item Catalogs}

\author{Gleb Mezentsev}
\orcid{0009-0003-7591-3082}
\affiliation{%
  \institution{Skolkovo Institute of Science and Technology}
  \city{Moscow}
  \country{Russian Federation}
}
\email{gleb.mezentsev@skoltech.ru}
\authornote{Both authors contributed equally to the paper}

\author{Danil Gusak}
\orcid{0009-0008-1238-6533}
\affiliation{%
  \institution{Skolkovo Institute of Science and Technology}
  \country{}
}
\affiliation{%
  \institution{HSE University}
  \city{Moscow}
  \country{Russian Federation}
}
\email{danil.gusak@skoltech.ru}
\authornotemark[1]

\author{Ivan Oseledets}
\orcid{0000-0003-2071-2163}
\affiliation{%
  \institution{Artificial Intelligence Research Institute}
  \country{}
}
\affiliation{%
  \institution{Skolkovo Institute of Science and Technology}
  \city{Moscow}
  \country{Russian Federation}
}
\email{oseledets@airi.net}

\author{Evgeny Frolov}
\orcid{0000-0003-3679-5311}
\affiliation{%
  \institution{Artificial Intelligence Research Institute}
  \country{}
}
\affiliation{%
  \institution{Skolkovo Institute of Science and Technology}
  \country{}
}
\affiliation{%
  \institution{HSE University}
  \city{Moscow}
  \country{Russian Federation}
}
\email{frolov@airi.net}


\begin{abstract}
  
Scalability issue plays a crucial role in productionizing modern recommender systems. Even lightweight architectures may suffer from high computational overload due to intermediate calculations, limiting their practicality in real-world applications. Specifically, applying full Cross-Entropy (CE) loss often yields state-of-the-art performance in terms of recommendations quality. Still, it suffers from excessive GPU memory utilization when dealing with large item catalogs.
This paper introduces a novel Scalable Cross-Entropy (SCE) loss function in the sequential learning setup. It approximates the CE loss for datasets with large-size catalogs, enhancing both time efficiency and memory usage without compromising recommendations quality.
Unlike traditional negative sampling methods, our approach utilizes a selective GPU-efficient computation strategy, focusing on the most informative elements of the catalog, particularly those most likely to be false positives.
This is achieved by approximating the softmax distribution over a subset of the model outputs through the maximum inner product search.
Experimental results on multiple datasets demonstrate the effectiveness of SCE in reducing peak memory usage by a factor of up to $100$ compared to the alternatives, retaining or even exceeding their metrics values.
The proposed approach also opens new perspectives for large-scale developments in different domains, such as large language models.

\end{abstract}

\begin{teaserfigure}
\vspace{-15pt}
\setlength{\abovecaptionskip}{4pt}
\centering
    \includegraphics[width=0.93\textwidth]{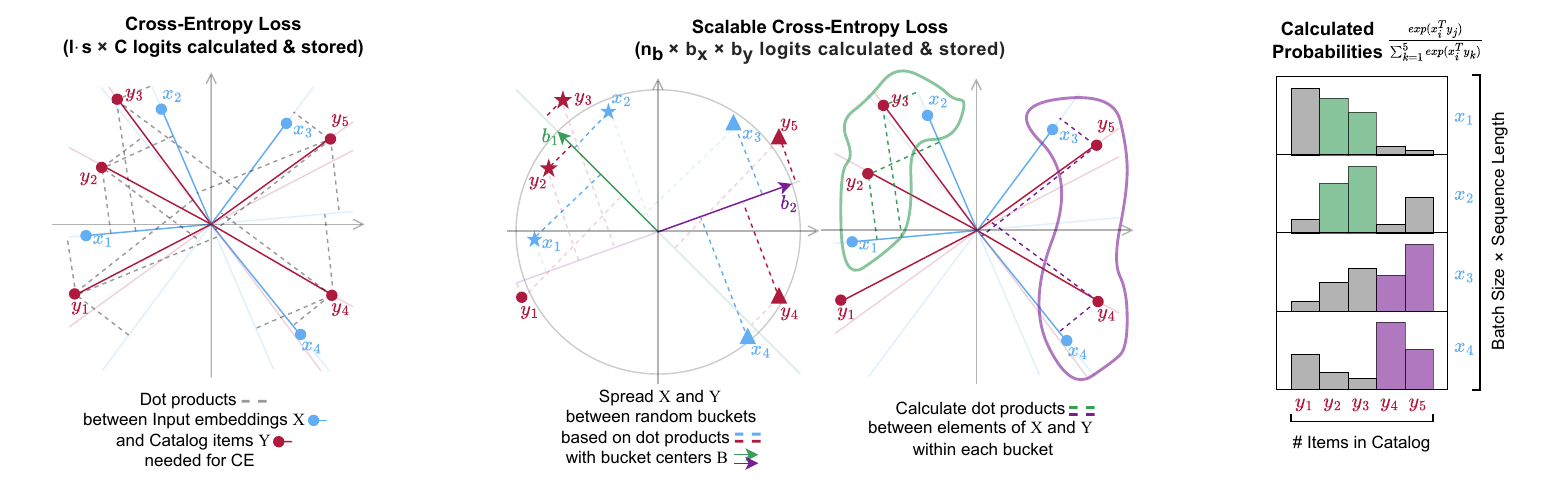}
    \caption{Schematic representation of logits computation for Cross-Entropy Loss and Scalable Cross-Entropy Loss with sequence length $l=4$, batch size $s=1$, catalog size $C=5$, number of buckets $n_b=2$ and bucket sizes $b_x=2$, $b_y=2$.}
    \label{fig:scheme}
\end{teaserfigure}




\maketitle

\section{Introduction}
\label{sec:intro}
\newlist{customitemize}{enumerate}{1}
\setlist[customitemize,1]{label=(\arabic*), leftmargin=*}

In collaborative filtering, many of the recent state-of-the-art models have increasingly adopted sequential approaches.
Sequential recommender systems aim to forecast the next item a user might choose based on past activity, a task known as \emph{next item prediction}. By considering the order of interactions, such systems can effectively suggest subsequent actions, such as recommending laptop accessories immediately after a user buys a laptop. This enhances the relevance and timeliness of recommendations.

\begin{figure}[t!]
\setlength{\abovecaptionskip}{-3pt}
\setlength{\belowcaptionskip}{-13pt}
    \centering
    \includegraphics[width=1.0\columnwidth]{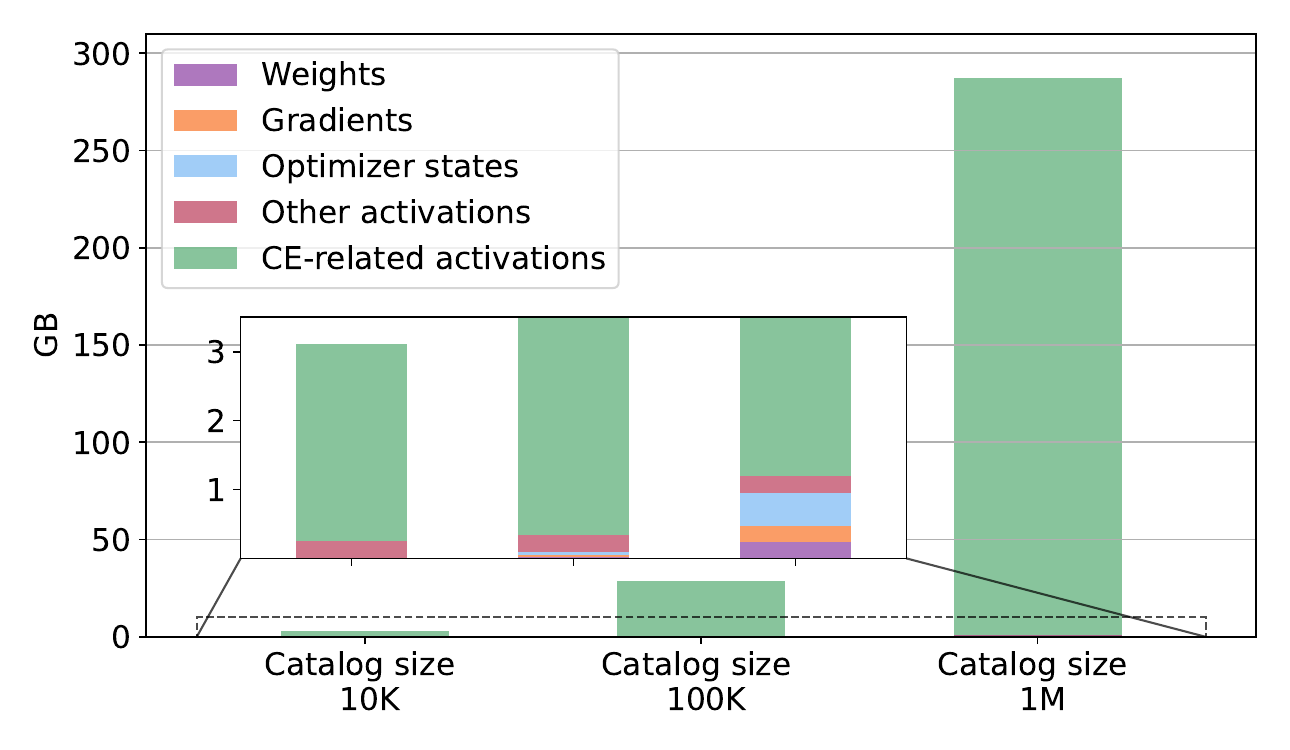}
    \caption{Impact of different components on peak GPU memory when training SASRec with Cross-Entropy loss. The measurements are performed using PyTorch library memory profiling tools.}
    \label{fig:memory}
    \Description[]{}
\end{figure}

Expanding beyond their original application in natural language processing (NLP), Transformer models \cite{vaswani2017attention} have been successfully leveraged to tackle the next item prediction task in recommender systems. 
Among the earliest notable works to accomplish this were SASRec \cite{kang2018self} and BERT4Rec \cite{sun2019bert4rec}, which were inspired by the success of GPT \cite{radford2018gpt} and BERT \cite{devlin2018bert} architectures, respectively. Despite many similarities to the original design, SASRec and BERT4Rec adapt certain modifications in their training objectives and attention mechanism. The authors of BERT4Rec attributed superior performance over SASRec to the bidirectional nature of the model. However, subsequent research \cite{Klenitskiy_2023, Petrov_2023} has revealed that the primary factor driving this improvement is the difference in the loss function used, while other modifications actually have a detrimental effect. SASRec was trained with Binary Cross-Entropy (BCE) loss \eqref{eq:bce_} with one positive class and one negative class, while BERT4Rec used the Cross-Entropy (CE) loss \eqref{eq:ce_} over the entire item catalog.

This highlights the superiority of the CE loss over its binary counterpart. However, the latter is often preferred in practice due to its better scalability on larger item catalogs. The problem with the full CE loss is the need to compute and store a \emph{tensor of logits} with $s \cdot l \cdot C$ elements corresponding to pairwise dot products of the outputs of the model (inputs to the classification head) and catalog embeddings, where $s$ is the training batch size, $l$ is the sequence length, and $C$ is the size of the catalog. Consider training on a batch of $128$ sequences, each of length $200$, with a catalog of $10^6$ items. It already requires about $100$GB just to store this tensor, which poses a challenge even for modern GPUs. Moreover, Transformers used in sequential recommendations are usually small (e.g., $2$ layers in SASRec vs $12$ layers in GPT), which makes the issue even more pronounced as almost all the memory during training is spent on the forward and backward passes through the loss function (Fig. \ref{fig:memory}).

This sets the stage for a compelling challenge: \emph{to develop a replacement for the CE loss that maintains a similar level of accuracy while operating within memory constraints akin to BCE}. The solution comes from the key difference between BCE and CE -- the former samples a single negative example, skewing the distribution of observed input data and impairing the learning ability, while the latter incorporates all available examples for the computation, which affects scalability. An ideal solution would be to select a subset of the most informative examples from the entire catalog that provide just enough information about input data without significant computational overhead. The most promising candidates for such selection are instances where the model misclassifies a negative example as positive, the so-called ``hard negatives''. 

Motivated by this intuition, we propose a novel Scalable Cross-Entropy (SCE) loss. \emph{It utilizes a selective computation strategy}, simultaneously prioritizing the most informative elements from input sequences and the item catalog -- exactly those that are most likely to cause errors. This is achieved by approximating the softmax distribution over a subset of the model's logits. These logits are pre-identified using a GPU-friendly approximate search for maximum inner products between two sets of embeddings, \emph{which eliminates the need to calculate all possible products}. SCE effectively reduces the size of the logit tensor by concentrating memory resources on the most informative cases, which mitigates inefficiencies associated with random or less selective negative sampling methods.

We further integrate SCE into the state-of-the-art sequential SASRec model for our experiments. 
It is worth noting that several recently published methods have partially or fully focused on solving the problem of large catalogs \cite{Petrov_2023, Klenitskiy_2023}. However, the results on memory reduction are rarely described in detail, although this would seem to be an important point for evaluating the proposed methods. In this paper, we consider the solution to the problem of extensive catalogs in the \emph{metric-memory trade-off paradigm}. We also demonstrate that the use of some negative sampling-based models is inappropriate for small catalogs, and in such cases, it is reasonable to use CE, both in terms of metrics and memory consumption.

The empirical evaluation of SASRec-SCE on five datasets (BeerAdvocate, Behance, Amazon Kindle Store, Yelp, and Gowalla) demonstrates that SCE is better than the baseline methods (BCE with multiple negatives, gBCE \cite{Petrov_2023}, sampled CE \cite{Klenitskiy_2023}), it retains the benefits of full CE loss in terms of achieved metrics, while significantly reducing the memory overhead. Furthermore, for the Amazon Beauty dataset, a comparison of SASRec-SCE to recently proposed models in the literature, which report the best results, showcases the performance of our method on par with SOTA models. While the primary focus of this paper is on sequential recommendation, the proposed method has the potential to be applied in other research domains, including natural language processing and search systems.

In short, the main contributions of this paper are as follows:
\aptLtoX{\begin{enumerate}
  \item We propose a memory-efficient SCE loss that can be applied to domains beyond sequential recommenders;
  \item We use SCE to train SASRec, perform an extensive evaluation of SASRec-SCE, and show that SCE allows to significantly reduce the peak training memory without compromising performance;
  \item We examine how a small catalog affects CE, SCE, and recent negative sampling methods, noting when each is more effective. 
\end{enumerate}}{\begin{customitemize}
  \item We propose a memory-efficient SCE loss that can be applied to domains beyond sequential recommenders;
  \item We use SCE to train SASRec, perform an extensive evaluation of SASRec-SCE, and show that SCE allows to significantly reduce the peak training memory without compromising performance;
  \item We examine how a small catalog affects CE, SCE, and recent negative sampling methods, noting when each is more effective. 
\end{customitemize}}

\section{Related Work}
\label{sec:literature}
In this section, we demonstrate the prevalence of Transformer-based models in the sequential recommendation setting and present a variety of methods that can be used for CE approximation or negative samples retrieval.

\subsection{Transformer-Based Sequential Recommenders}\label{sec:seqrec}

Since the introduction of the Transformer architecture \cite{vaswani2017attention}, Trans\-former-based models have consistently outperformed other approaches in the area of sequential recommendations \cite{kang2018self, sun2019bert4rec, Xie2022ContrastiveLF, petrov2023generative}. Following the original SASRec \cite{kang2018self} and BERT4Rec \cite{sun2019bert4rec} papers, extensive research was conducted to investigate the potential of Transformer-based models further. Several papers have introduced contrastive learning for sequential recommendations (DuoRec \cite{duo}, CBiT \cite{du2022contrastive}, CoSeRec \cite{cosorec}). Some works have tried to use additional side information (NOVA-BERT \cite{liu2021noninvasive}, TiSASRec \cite{tisas}). Others have focused on improving the self-attention mechanism (Rec-Denoiser \cite{denoise}, LightSAN \cite{light}). 

As we discuss in Section \ref{sec:intro}, SASRec exhibits state-of-the-art performance when trained with CE (\ref{eq:ce_}). However, the large size of the item catalog in many real-world applications means that the use of negative sampling methods, such as BCE (\ref{eq:bce_}), or other approximation strategies is unavoidable when training such systems. Our goal is, therefore, to maximise the performance of SASRec while maintaining its scalability to large catalogs.

\begin{equation}
    \mathcal{L}_{CE} = -\log \frac{\exp(logit_{i, +})}{\sum_{c=1}^C \exp(logit_{i, c})}
    \label{eq:ce_}
\end{equation}

\begin{equation}
    \mathcal{L}_{BCE} = - \log (\sigma(logit_{i, +})) - \log (1 - \sigma(logit_{i, -}))
    \label{eq:bce_}
\end{equation}

\subsection{Approaches to Negative Sampling and Cross-Entropy Approximation}\label{sec:sampling}

One of the most straightforward methods to sample negative examples is uniform random sampling \cite{tang2018personalized, kang2018self}, where one or several negative samples are drawn from the set of negative items uniformly. The performance could be improved by increasing the number of negative samples \cite{Klenitskiy_2023}. This could be achieved by making slight modifications to either the BCE loss function, as in Caser model \cite{tang2018personalized}, or the CE loss function, as in \citet{Klenitskiy_2023}. The modifications are outlined below:
\begin{equation}
    \mathcal{L}_{BCE^+} = - \log (\sigma(logit_{i, +})) - \sum_{j \in I_k^-} \log (1 - \sigma(logit_{i, j})),
    \label{eq:bce_2}
\end{equation}
\begin{equation}
    \mathcal{L}_{CE^-} = -\log \frac{\exp(logit_{i, +})}{ \exp(logit_{i, +}) + \sum_{j \in I_k^-} \exp(logit_{i, j})},
    \label{eq:ce_2}
\end{equation}
where $I_k^-$ is a set of $k$ sampled negatives.

However, the performance suffers from completely random nature of this procedure. One of the consequences is that the model rarely sees hard negatives and becomes overconfident -- it predicts very high probabilities for all high-ranked items and is not able to identify the one true item \cite{Petrov_2023}. The issue can be addressed by calibrating the predicted scores using the information about the item catalog \cite{Petrov_2023}. This adjustment allows us to fight the overconfidence problem, however, the negative samples still remain uninformative. Another popular solution is popularity-based sampling. Although it often yields better results than uniform sampling \cite{lian2020personalized, chen2022generating}, it is shown to be outperformed by other methods, that try to target hard negatives directly \cite{chen2022generating, rendle2014improving}. It also gives the most benefits when popularity-based metrics are used for evaluation \cite{pellegrini2022don}, which lately were shown to be problematic \cite{Dallmann_2021, rocio21, 50163}. In-batch negative sampling \cite{hidasi2015session, hidasi2018recurrent} is a related method, which uses true class labels for other items in a batch as negative samples. It is also based on item popularity, as more popular items are more likely to be present as positive class labels.

The other possible approach, which is targeting more informative samples, is to employ an approximation of the softmax distribution to sample from. Possible variants for approximation include matrix factorizations \cite{rendle2014improving}, fixed \cite{bengio2003quick} and adaptive n-grams \cite{bengio2008adaptive} and kernel methods \cite{blanc2018adaptive, rawat2019sampled}. A popular alternative to such sampling is a two-step procedure with sampling of a large set of items uniformly followed by selecting a smaller subset of items with larger logit values that will be used for loss computation \cite{bai2017tapas, chen2022generating, wilm2023scaling}.

Finally, there is a group of methods that aim to select useful (hard) negative samples for loss computation. Several authors suggested accumulating hard negatives for each user between training iterations \cite{wang2021cross, ding2020simplify}, with the reasoning that item representations do not change too quickly and hard negatives could be reused. However, this method introduces an obvious overhead of maintaining an additional data structure in memory.

This overhead can be avoided if hard negatives are selected at each step. Since hard negatives are just items with large logits (large inner products with the current embedding), the problem of searching for hard negatives could be substituted with (Approximate) Maximum Inner Product Search (MIPS) or (Approximate) Nearest Neighbour Search (NNS), which was done in several works. Some of them rely on various Locality Sensitive Hashing (LSH) schemes to build hash tables and use them to retrieve relevant items \cite{vijayanarasimhan2014deep, spring2017new, guo2016quantization}, while others \cite{yen2018loss, lian2020personalized} use variations of spherical clustering \cite{auvolat2015clustering} as a tool for solving MIPS, or utilize \cite{mussmann2016learning, mussmann2017fast} more general MIPS schemes combining them with Gumbel sampling \cite{gumbel1954statistical}. The common issue with these methods and all existing (to the best of our knowledge) MIPS and NNS methods, is their poor compatibility with GPU computations. It is very desirable for the efficient GPU utilization to perform operations in a batch manner. This is nontrivial in the case of the methods mentioned above. This claim is supported by the fact, that none of these works presented experiments conducted on GPU, even though GPU training was already dominant when those works were published. As the limiting factor, preventing efficient and easy GPU implementations of these methods, we see the inability to guarantee the same size of the set of retrieved samples for each item in a batch. This is the key difference with our method, which allows us to fully utilize batching while processing the data.

We can see that existing methods either do not target hard negatives or are inefficient for GPU computations which leads to suboptimal quality or limited applicability of the resulting model.

\section{Proposed approach}
\label{sec:approach}
In this section, we propose a novel scalable approach to approximate Cross-Entropy loss that significantly reduces memory requirements while maintaining performance. Furthermore, we describe a potential problem called "bucket collapse" that may limit the effectiveness of our approach, examine a strategy to mitigate this problem and discuss the broad applicability of our approach across different domains.

\subsection{Scalable Cross-Entropy}
\label{sec:main_method}

The goal of our method is to replace the calculation of the Cross-Entropy loss function over the entire catalog with the calculation of the same function over the part of the catalog that affects the gradient update the most and to find that part in a GPU-friendly manner.

Suppose we want to make a prediction of the next item for item $z_i$ (catalog index), the correct answer for which is $z_{i+1}$. Let the output for item $z_i$ of our transformer (before the classification head) be $x_i$, and the current embedding of item $j$ be $y_j$ (both $x_i$ and $y_j$ of dimension $d$). If the size of the catalog is $C$, then the value of Cross-Entropy loss for item $z_i$ is computed as follows:
\begin{equation}
    \begin{split}
    \mathcal{L}_{CE} = \mathcal{L}_{CE}(x_i, z_{i+1}) = -\log(\text{softmax}(logit_i)_{z_{i+1}})\ = \\
    = -\log \frac{\exp(logit_{i, z_{i+1}})}{\sum_{c=1}^C \exp(logit_{i, c})} = - \log\frac{\exp(x_i^Ty_{z_{i+1}})}{\sum_{c=1}^C \exp(x_i^Ty_c)}
    \end{split}
    \label{eq:ce}
\end{equation}
\begin{equation}
    \begin{split}
    \frac{\partial\mathcal{L}_{CE}}{\partial logit_{i,k}}  = \text{softmax}(logit_i)_k-\mathds{1}[k=z_{i+1}]
    \end{split}
    \label{eq:ce_grad}
\end{equation}

It can be seen from Eq. \ref{eq:ce_grad} that the gradient of the Cross-Entropy loss function with respect to logits takes values between $-1$ and $1$. It is close to $1$ if the predicted probability is close to $1$ while $k$ is not a correct class (we predicted a high probability for the wrong class), and it is close to $-1$ if the predicted probability is close to $0$, while $k$ is a correct class (we predicted a low probability for the correct class).

We want to compute only logits with the largest absolute values of the gradients (maximize the amount of preserved information after our approximation), so we need to find such cases in advance. The instances of the second type are easy to find because we know the only correct class for each input item. The instances of the first type are harder to find, and we substitute the problem of finding large logits of negative classes with the problem of finding any large logits (because, for any item, it is at most one logit of positive class in the set of large logits), which is essentially a MIPS problem.

To solve this problem, we introduce a set of random vectors -- $B\in \mathbb{R} ^{n_b\times d}$, bucket centers, such that $|B| = n_b << C$ and $|B|= n_b << l \cdot s$, where $s$ and $l$ are input batch size and sequence length. For each vector from $B$, we build a neighborhood (bucket) around it -- select top-k closest (with the highest dot product) vectors out of model outputs $X$ and catalog embeddings $Y$, and calculate logits only between vectors within the bucket.
The intuition behind this idea is as follows: if we have two vectors $x_i$ and $y_j$ that are both close (have a large dot product) to some random vector $b$, then $x_i$ and $y_j$ are also probably close. This splitting procedure is cheap as it is based on GPU-efficient matrix multiplications. Moreover, the buckets constructed based on top-k selection are of the same size, which enables efficient batched computations. 

This idea is similar to Spherical Clustering \cite{auvolat2015clustering}, which is a possible approach for MIPS in the context of Softmax Approximation \cite{yen2018loss, lian2020personalized}, however, the key difference is that our method allows us to construct buckets of the same size, and as a result perform computations on GPU efficiently via batching. 

Another similar approach is the adaptation of the so-called Locality Sensitive Hashing for angular distance \cite{andoni2015practical} for Cross-Entropy loss approximation, a method called Reduced Cross-Entropy loss (RECE) \cite{gusak2024rece}. Then main difference is that our method utilizes a different approach for spreading the objects between the buckets, which provides two benefits. We have more control for memory consumption by having the bucket size as a hyperparameter in contrast to RECE, where all input objects are split between the buckets, so the bucket size cannot be changed and is determined by the number of buckets. We also increase the efficiency of the computations -- our buckets have the same size by design, so we don't have to do the chunking step, introduced in \citet{gusak2024rece}, where neighboring buckets are mixed and split into new buckets of the same size, where not all objects will be relevant. These benefits result in significant improvement in the performance, which is shown in Table \ref{tab:recentresults}.

Let's look at our Algorithm \ref{alg:rce_new} in more detail. The first step (Line 2) is the generation of a set of random vectors $B$, which is a matrix with values sampled from $\mathcal{N}(0, 1)$. The next step (Lines 3-11) is projecting $X$ and $Y$ onto generated vectors $b$ and finding $b_x$ vectors from $X$ and $b_y$ vectors from $Y$ with the largest dot products for each of these random vectors. This gives the distribution of the vectors between the buckets. This distribution could be overlapping (some of the vectors could end up in multiple buckets), but it guarantees buckets of the same size with no completely irrelevant vectors inside buckets (in contrast to RECE~\cite{gusak2024rece}). After that, the logits for the wrong classes and the correct class are calculated (Lines 12-14). $\hat{Z}$ represents the set of correct predictions -- model inputs shifted to the right. It is important to note that in Line 12, some of the calculated logits could also correspond to the correct class, so these values should be masked (filled with $-\infty$) to block the passage of the gradients. The final step is the calculation of the Cross-Entropy loss values inside each bucket (Line 15) and the accumulation of these values between the buckets (Lines 16-17) -- some of the model outputs could be placed into several buckets, so we take the maximum value between the calculated values, as the maximum value corresponds to a partial sum over the catalog which is closer to the full sum.

The main memory bottleneck of full Cross-Entropy loss calculation is the storage of the logit tensor, which has a shape equal to $s\cdot l \times C$. In the case of our algorithm, the largest tensor to be stored (with the reasonable choice of hyperparameters) is also a tensor of logits, which has a shape equal to $n_b \times b_x \times b_y$, which could be significantly smaller.

This method has one potential source of inefficiency: if one input is placed into several buckets, then all but one calculated Cross-Entropy values are not used, due to the $max$ aggregation. So, to maximize efficiency, we want to minimize the situations where different buckets share the same objects (network outputs). The following section presents an approach to mitigate this problem.

\begin{algorithm}
   \caption{Scalable Cross-Entropy Loss}
   \label{alg:rce_new}
\begin{algorithmic}[1]
\small
   \State {\bfseries Input:} $\hat{Z} \in \mathbb{N} ^{s \cdot l}$, $X \in \mathbb{R} ^{s \cdot l \times d}$, $Y \in \mathbb{R} ^{C \times d}$, $n_b$, $b_x$, $b_y$
   \State $B = \text{randn}(n_b, d) = [b_1, b_2, \dotsb , b_{n_b}]$
   \State {\bfseries With no gradient tracking:}
   \State\hspace{\algorithmicindent} $X^P = B X^T$
   \State\hspace{\algorithmicindent} $Y^P = B Y^T$
   \State\hspace{\algorithmicindent}
   {\bfseries for $b$ in $B$:}
    \State\hspace{\algorithmicindent}\hspace{\algorithmicindent} $I_b = \text{argsort} (X^P[b]) [-b_x:]$ \footnotesize\Comment{select top-$b_x$}\small
    
    \State\hspace{\algorithmicindent}\hspace{\algorithmicindent} $J_b = \text{argsort} (Y^P[b]) [-b_y:]$ \footnotesize\Comment{select top-$b_y$}\small
    
    \State
    {\bfseries for $b$ in $B$:}
   
   \State\hspace{\algorithmicindent} $B^X_{b} = X[I_{b}]$
   \State\hspace{\algorithmicindent} $B^Y_{b} = Y[J_{b}]$

   \State\hspace{\algorithmicindent} $logits^-_b = B^X_b {B_b^Y}^T  \in \mathbb{R} ^{b_x\times b_y}$ \footnotesize\Comment{wrong class logits}\small

    \State\hspace{\algorithmicindent}
    {\bfseries for $i$ in range($b_x$):} 
    \State\hspace{\algorithmicindent}\hspace{\algorithmicindent} $logit^+_{b, i} = \sum_{m=1}^{d} X[I_{b, i}]_m \cdot Y[\hat{Z}[I_{b, i}]]_m$  \footnotesize\Comment{correct class logit}\small
    \State\hspace{\algorithmicindent}\hspace{\algorithmicindent} $loss_{b,i} = -\log \frac{\exp(logit^+_{b, i})}{\exp(logit^+_{b, i})+\sum_{j=1}^{b_y}\exp(logit^-_{b, i, j})}$
   
   \State $L_k = \{b,i: I_{b,i}=k\}$ \footnotesize\Comment{set of positions (bucket number, index inside bucket), where $x_k$ was placed}\small
   \State $\mathcal{L}_{SCE} = \frac{1}{|k: L_k \neq \varnothing|} \sum_{k: L_k \neq \varnothing} \max_{b,i \in L_k} loss_{b,i}$ \footnotesize\Comment{for every $x_k$ placed at least in one bucket, the maximum loss value over buckets is selected}\small
\end{algorithmic}
\end{algorithm}

\subsection{Bucket Collapse Mitigation}\label{sec:basis_generation}

In order to mitigate the problem of bucket collapse, several alternative approaches to the identification of similar vectors can be considered, including trainable clustering methods with fast searching, or methods based on information gain metrics. However, in this work, we focus on and analyze randomized approaches with random generation of $B$, favoring speed and reduced memory complexity. One advantage of such an approach is that, for each subsequent batch, a new random matrix is generated, which serves as a form of regularization. As previously stated, our goal is to minimize situations where different buckets share the same model outputs. The problem then reduces to finding a fixed number ($n_b$) of vectors, denoted as $B$, that optimally separate our model outputs in the computational space.

We proceed to resolve this issue by following \citet{halko2010finding} and multiplying a randomly generated matrix by the model output matrix, resulting in the operation designated as \textit{Mix}.
This way, we sample from $\mathcal{N}(0, 1)$ a matrix $\Omega \in \mathbb{R}^{n_b \times s \cdot l}$, consisting of $n_b$ vectors, then project this matrix onto model outputs $B = \Omega X \in \mathbb{R}^{n_b \times d}$. The resulting matrix, $B$, is then processed, continuing from Algorithm \ref{alg:rce_new}, Line 3. The intuition behind this operation is that it allows us to identify informative directions in space -- the projection of the vectors onto those directions will be large in magnitude because the directions are constructed as a combination of these vectors. At the same time, due to the random nature of the operation, these directions will be sufficiently diverse, which will help to cover more space compared to a deterministic selection. In comparison to the initial method, an additional matrix of size $n_b \times s \cdot l$ is stored, which is negligible in the case of large catalogs.
It should be noted that instead of $X$, \textit{Mix} can also be applied to Y, a matrix of catalog embeddings. However, our study demonstrated that this operation does not have a positive effect on performance.  Therefore, in our experiments, we apply Mix with respect to the matrix of model outputs $X$. The empirical evaluation demonstrates that this simple operation, performed for each batch and requiring an insignificant amount of memory and computation, allows for a notable improvement in model quality metrics.

\subsection{Method Applicability}\label{sec:versatility}

In this work, we utilize SASRec as our "backbone" model to test our approaches, given its status as a widely used architecture in the literature \cite{Klenitskiy_2023, cls4rec, Petrov_2023} that exhibits state-of-the-art performance in sequential recommendations when trained with full Cross-Entropy loss \cite{Klenitskiy_2023} (see Sections \ref{sec:intro}, \ref{sec:seqrec}).

 Although our primary focus on the SCE method is on recommender systems, where managing large catalogs is a common problem, the applicability and potential of this approach extend across various domains. In natural language processing, the diversity of words in a text corpus increases with its size \cite{dodge2021documenting}. Large vocabulary presents a significant challenge to the effective computation of logits across all tokens, and our method represents a viable contribution to the solution of this challenge. For example, GPT-like architectures \cite{Radford2019LanguageMA} with extensive vocabularies could benefit from the efficient computation of Cross-Entropy loss. Recently, large language models (LLMs) (based on such architectures) have been used for sequence prediction \cite{xue2023promptcast}, which provides another possible application of our method for recommender systems.  The method can also be applied to problems in computer vision, bioinformatics \cite{nbio}, search systems, and other areas, where tasks with a large number of classes or pairwise interactions are common, showcasing its versatility and potential for enhancing training memory efficiency and model performance.

\section{Experiments}
\label{sec:experiments}

In this section, we describe the design and results of the experiments that explore SCE's applicability and competitive advantage.

\subsection{Experimental Settings}
\label{sec:experimental_settings}

\subsubsection{Datasets}\label{sec:datasets}

We conduct our main experiments on five datasets collected from real-world platforms, which vary significantly in data statistics and domains:

\begin{itemize}[leftmargin=*] 
\item  \textbf{BeerAdvocate \cite{mcauley2012beer}:} this dataset consists of beer reviews from BeerAdvocate.

\item  \textbf{Behance \cite{behance}:} the dataset comprises likes and image data from the community art website Behance. It is a smaller, anonymized version of a larger proprietary dataset.

\item  \textbf{Kindle Store \cite{kindle}:} collection of data extracted from Amazon's Kindle Store, which is a section of Amazon's website dedicated to the sale of e-books, e-magazines, and other digital reading content.

\item  \textbf{Yelp \cite{asghar2016yelp}:} business reviews dataset \cite{padungkiatwattana2022yelp, amjadi2021yelp, ruihong2021yelp}.

\item  \textbf{Gowalla \cite{cho2011gowalla}:} originating from the location-based social networking platform Gowalla, the dataset includes user check-ins.

\end{itemize}

Consistent with prior research \cite{tang2018personalized, kang2018self, sun2019bert4rec}, the presence of a review or rating is interpreted as implicit feedback. Furthermore, following common practice \cite{kang2018self, rendle2010factorizing, zhang2019feature} and to ensure the number of items in datasets allows for the computation of full Cross-Entropy loss within GPU memory constraints, we exclude unpopular items with fewer than $5$ interactions and remove users who have fewer than $20$ interaction records. The final statistics of the datasets are summarized in Table \ref{tab:datasetStats}. As demonstrated, the number of items across these datasets ranges from a comparatively small $22,307$  in BeerAdvocate, to a large $173,511$ in Gowalla, facilitating the evaluation of our approach and baselines under various memory consumption conditions.

\begin{table}[]
\setlength{\abovecaptionskip}{3pt}
\setlength{\belowcaptionskip}{-1pt}
\footnotesize
\caption{\textbf{{Statistics of the datasets after preprocessing.}}} \label{tab:datasetStats}
\resizebox{0.9\columnwidth}{!}{%
\begin{tabular}{lrrrr}

\hline
Dataset      & Users  & Items   & Interactions & Density \\ \hline
BeerAdvocate & 7,606  & 22,307  & 1,409,494    & 0.83\%  \\
Behance  & 8,097  & 32,434  & 546,284      & 0.21\%  \\
Kindle Store & 23,684 & 96,830  & 1,256,065    & 0.05\%  \\
Yelp         & 42,429 & 137,039 & 2,294,804    & 0.04\%  \\
Gowalla      & 27,516 & 173,511 & 2,627,663    & 0.06\%  \\ \hline
\end{tabular}%
}
\end{table}

\subsubsection{Evaluation}\label{sec:evals}

In offline testing, data splitting using the leave-one-out approach, which usually involves the selection of the last interaction for each user for testing, is a common practice mentioned in earlier studies \cite{meng2020exploring, Petrov_2023, Klenitskiy_2023}. Although this method is widely used, there have been concerns about data leakages, which affect the accuracy of the evaluation results for recommendation models \cite{Ji_2023}.
To address this issue, we set a global timestamp that corresponds to the $0.95$ quantile of all interactions \cite{frolov2022tensorbased}. Interactions before the timestamp are used for training. Users who interact after this point are considered for testing, keeping them separate from the training data. For these test users, we follow the standard leave-one-out procedure, using their latest interaction for the evaluation of model performance. By excluding these users from the training dataset and filtering it by timestamp, we ensure the elimination of “recommendations from future” bias \cite{meng2020exploring}. This approach guarantees that the model remains uninformed about subsequent interaction patterns, thereby addressing the potential for temporal data leakage. Additionally, for each test user, we choose their second-to-last interaction for a validation set, which we use for hyperparameter tuning and to control model convergence through the early stopping mechanism.

In recent studies, the evaluation of recommender systems has shifted from relying on sampled metrics, where a limited number of negative items are compared against each positive item, to full unsampled metrics. This transition is driven by evidence suggesting that sampled metrics can lead to inconsistent and potentially misleading performance evaluations \cite{Dallmann_2021, rocio21, 50163}. To align with established best practices, our analysis employs widely accepted unsampled top-K ranking metrics: Normalized Discounted Cumulative Gain (NDCG@K) and Hit Rate (HR@K), with K $= 1, 5, 10$. Since our primary objective examines the balance between time consumption, memory efficiency of the model, and its ranking performance, we also measure training time and peak GPU memory during the training phase. Additionally, we incorporate Coverage (COV@K) as a metric for evaluating the diversity of the recommendations.

\subsubsection{Model and Baselines}\label{sec:baselines}

As previously stated in Section \ref{sec:versatility}, in our experiments, we utilize SASRec as the "backbone" model and enhance it with the proposed SCE loss. Our focus is on comparing \textbf{SASRec-SCE} with the model employing Binary Cross-Entropy loss with multiple negative samples (\textbf{SASRec-BCE$^+$}), detailed in Equation (\ref{eq:bce_2}), and with SASRec incorporating full Cross-Entropy loss (\textbf{SASRec-CE}).
Furthermore, we explore variations of the loss function for SASRec proposed by \citet{Klenitskiy_2023} and \citet{Petrov_2023}, which represent the most recent and relevant baselines. These are referred to as \textbf{SASRec-CE$^{-}$} and \textbf{gSASRec} (with gBCE), respectively. 
To ensure a consistent framework across our experiments, we base all the above models on the adapted PyTorch implementation\footnote{\url{https://github.com/pmixer/SASRec.pytorch}} of the original SASRec model architecture, and augment them with loss functions and sampling strategies described in the corresponding papers. The code is available in our GitHub repository\footnote{\url{https://github.com/AIRI-Institute/Scalable-SASRec}\label{github}}.

\subsection{Results}\label{sec:results0}

\subsubsection{Dependence on SCE Hyperparameters}\label{sec:results_hyperparams}

SASRec-SCE model has three model-specific hyperparameters: $n_b$, $b_x$ and $b_y$. As $n_b$ and $b_x$ control the number of model outputs that are processed and their distribution between buckets, we considered the following parametrization of these parameters: $b_x = \alpha \sqrt{s \cdot \overline{l} \cdot \beta}$, $n_b = \alpha \sqrt{s \cdot l / \beta}$, where $\overline{l}$ is an average number of interactions per user in a given dataset. In this parametrization, $\beta = n_b / b_x$ allows selecting between a larger number of smaller buckets and a smaller number of larger buckets, and $\alpha = \sqrt{(n_b \cdot b_x) / (s \cdot \overline{l})}$ acts as an oversampling coefficient. According to our preliminary experiments, the optimal value for $b_y$ depends on the dataset and $s$ selected and needs to be tuned depending on the memory budget.

\begin{figure}
\setlength{\abovecaptionskip}{6pt}
    \includegraphics[width=\columnwidth]{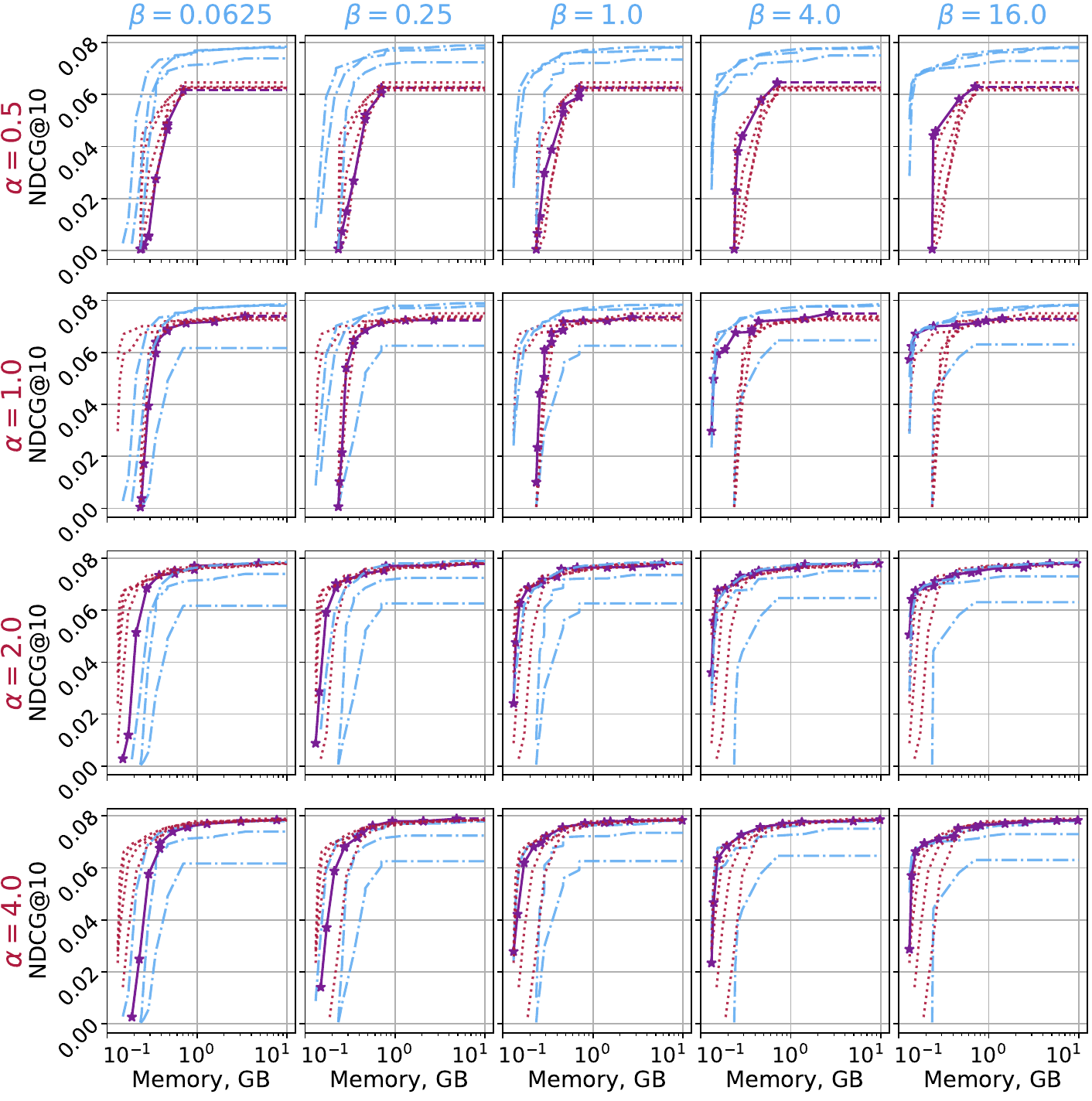}
    \caption{Effect of $\alpha$ and $\beta$ on SASRec-SCE performance on Kindle Store dataset. Each curve is a Pareto front for different values of $s$ and $b_y$. Curves corresponding to a $(\alpha, \beta)$ pair are in purple solid lines, curves corresponding to the same $\alpha$ are red dotted lines, and curves corresponding to the same $\beta$ are dash-dotted blue lines. Dashed lines indicate that no configurations yield a higher NDCG@10 for a larger GPU memory budget.}
    \label{fig:hyperparams}
\end{figure}

Figure \ref{fig:hyperparams} demonstrates how different values of $\alpha$ and $\beta$ affect model quality on the Kindle Store dataset. For each value pair $(\alpha, \beta)$, we probed a grid of values for $s$ and $b_y$ and constructed the Pareto front\footnote{\url{https://en.wikipedia.org/wiki/Pareto_front}} (a set of optimal configurations) in terms of achieved NDCG@10 for a given memory budget. All combinations of $\alpha \in \{2, 4\}$ and $\beta \in \{1, 4, 16\}$  (lower right portion of the graph) yield approximately the same optimal quality, so for later experiments, we fixed $\alpha = 2$ and $\beta = 1$.

\subsubsection{Influence of Mix Operation}\label{sec:results1}

\begin{figure*}[]
    \centering
    \setlength{\abovecaptionskip}{0pt} 
    \setlength{\belowcaptionskip}{-5pt} 

    \begin{minipage}{0.49\textwidth}
    \setlength{\abovecaptionskip}{0pt} 
    \setlength{\belowcaptionskip}{0pt} 
        \centering
        \includegraphics[width=\textwidth]{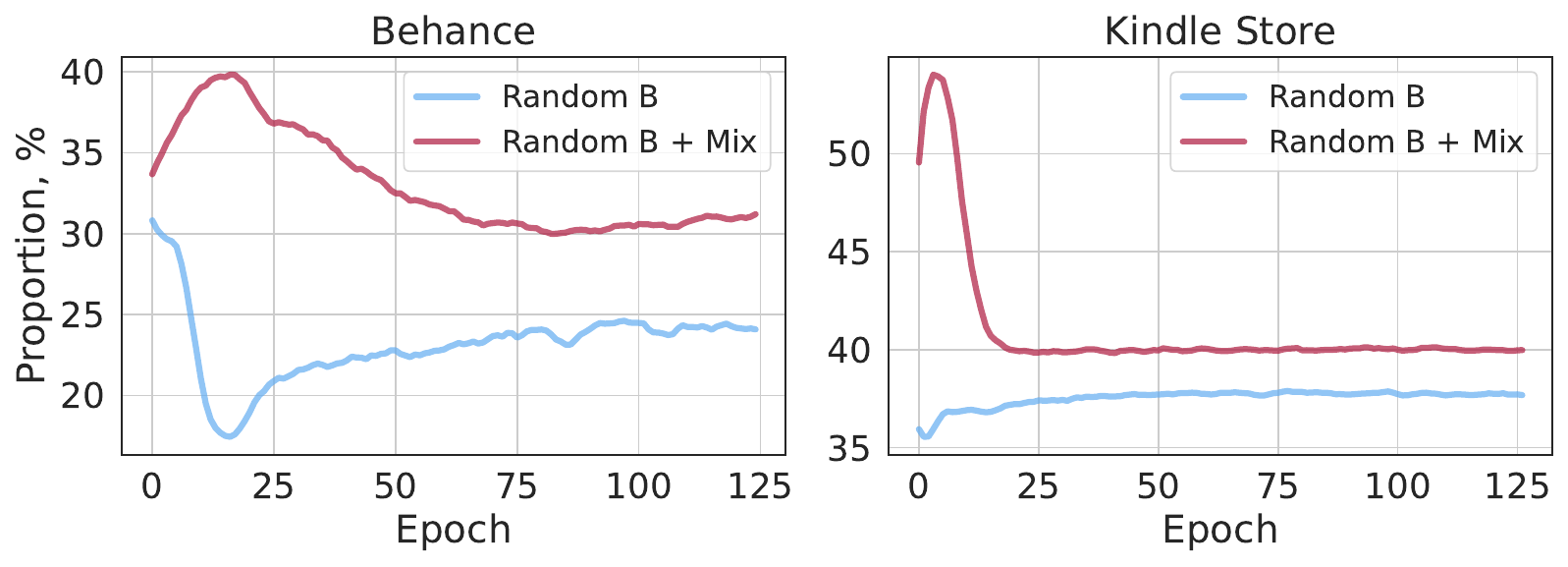}
        \subcaption{}
        \label{fig:part_unique_plot}
    \end{minipage}
    \hfill
    \begin{minipage}{0.49\textwidth}
    \setlength{\abovecaptionskip}{0pt} 
    \setlength{\belowcaptionskip}{0pt} 
        \centering
        \includegraphics[width=\textwidth]{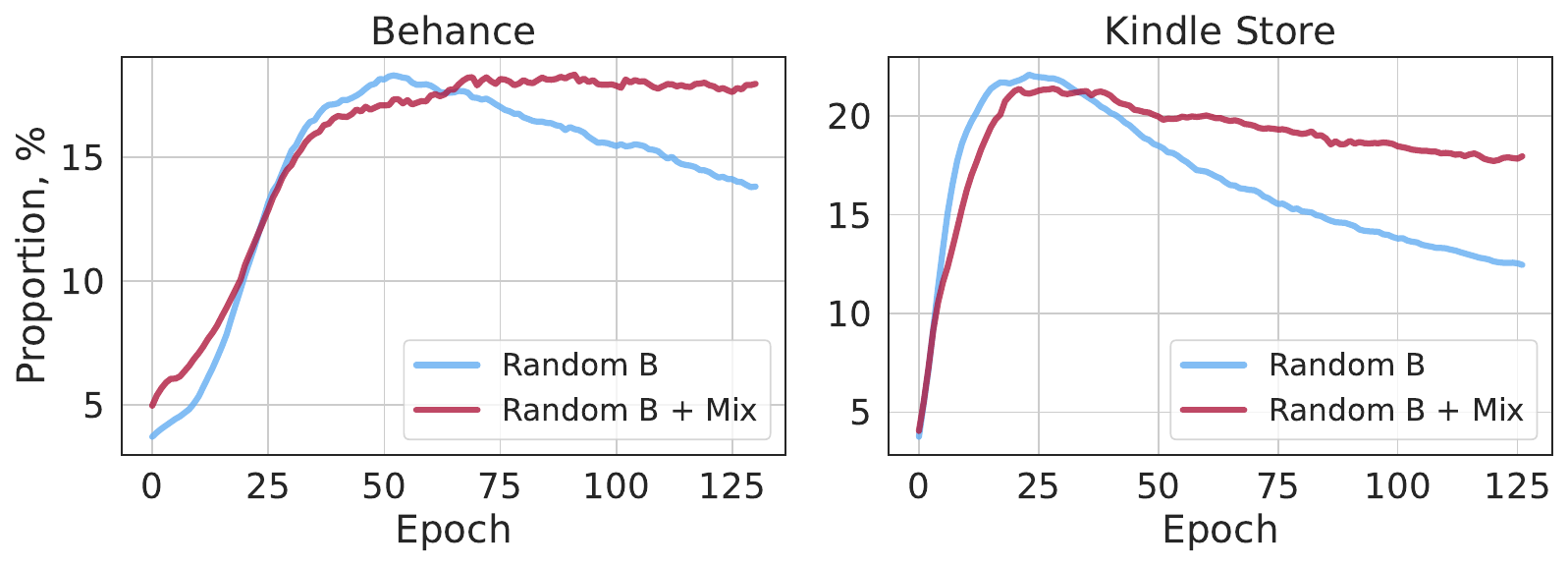}
        \subcaption{}
        \label{fig:part_logits_plot}
    \end{minipage}
    \caption{(a) Training dynamics of the proportion of model outputs of SASRec-SCE, which are selected only once within all buckets. (b) Training dynamics of the proportion of predicted logits for correct classes (items) in total number $n_b \cdot b_x$ of possible correct classes. Comparison of the construction methods for $B$, with and without the use of Mix.}
\end{figure*}

Following Section \ref{sec:basis_generation}, we evaluate the ability of generated buckets to separate model outputs $X$. This is achieved by observing how frequently elements of $X$ occur uniquely within all buckets.
In Figure \ref{fig:part_unique_plot} we compare two methods of constructing $B$: with and without the Mix operation. The figure illustrates the change in the in-batch average number of embeddings from $X$ that are selected only once within buckets $B$ during each step of the training phase. In this context, selection is defined as the process of selecting the top $b_x$ elements closest to $b$ in terms of the scalar product, described in Algorithm \ref{alg:rce_new}, Lines 7 and 10. The figure indicates that the proportion of unique selections is significantly higher when using Mix compared to the base method without Mix (at peak, 39.8\% vs 17.4\% and 54.0\% vs 37.1\%, for Behance and Kindle Store, respectively). This suggests that $B$ is better at distinguishing between different model outputs when constructed with Mix. This implies that the number of times the same model output is evaluated for different buckets is reduced, allowing a greater number of informative model outputs to be included in the final loss (Algorithm \ref{alg:rce_new}, Lines 16-17).

In addition, we evaluate the model's performance in terms of classification task. While this evaluation does not directly reflect the model's effectiveness in ranking- or relevance-based metrics, it provides valuable and interpretable insight into how accurately the model predicts correct classes. Following the SCE approach, for each batch, the total number of $n_b \cdot b_x \cdot b_y$ "wrong" class logits is computed (as specified in Algorithm \ref{alg:rce_new}, Line 12). Of these, at most $n_b \cdot b_x$ logits can correspond to scores of correct classes. We then examine the number of calculated logits for correct classes experimentally by analyzing the behavior of the models during training.
Figure \ref{fig:part_logits_plot} displays the training dynamics of the proportion of predicted logits for correct classes relative to the total number $n_b \cdot b_x$ of possible correct classes, comparing different methods of constructing $B$. As can be seen from the figure, the addition of the Mix operation to the SCE pipeline results in more stable and efficient learning. 

Furthermore, Table \ref{tab:ablation} summarizes the results of an ablation study. As can be seen from the table, Mix operation significantly improves the effectiveness of the model. For example, the NDCG@10 on the Kindle Store dataset improved from 0.0748 to 0.0780 (+4.3\%), and the improvement in COV@10 is 10.9\% (from 0.214 to 0.237).

\begin{table}[h]
\setlength{\abovecaptionskip}{3pt}
\caption{\textbf{{Effect of Mix operation on SASRec-SCE performance.}}} \label{tab:ablation}
\small
\resizebox{1.0\columnwidth}{!}{
\begin{tabular}{lcccccc}
\hline
\multirow{2}{*}{Dataset} & \multicolumn{2}{c}{NDCG@10} & \multicolumn{2}{c}{HR@10} & \multicolumn{2}{c}{COV@10} \\ \cline{2-7} 
             & w/o Mix & Mix    & w/o Mix & Mix    & w/o Mix & Mix     \\ \hline
Behance  & 0.0632  & +5.5\% & 0.1110  & +5.0\% & 0.215   & +17.8\% \\
Kindle Store & 0.0748  & +4.3\% & 0.1012  & +3.1\% & 0.214   & +10.9\% \\ \hline
\end{tabular}%
}
\end{table}

Thus, the incorporation of Mix, which requires a negligible amount of resources (see Section \ref{sec:basis_generation}), enhances the model's separating ability, leading to more stable and efficient training, which is supported by an increase in metrics.

\subsubsection{Small Catalog Effects}\label{sec:results2}

\begin{figure}[h]
    \setlength{\abovecaptionskip}{2pt} 
    \setlength{\belowcaptionskip}{-5pt} 
    \includegraphics[width=1.0\columnwidth]{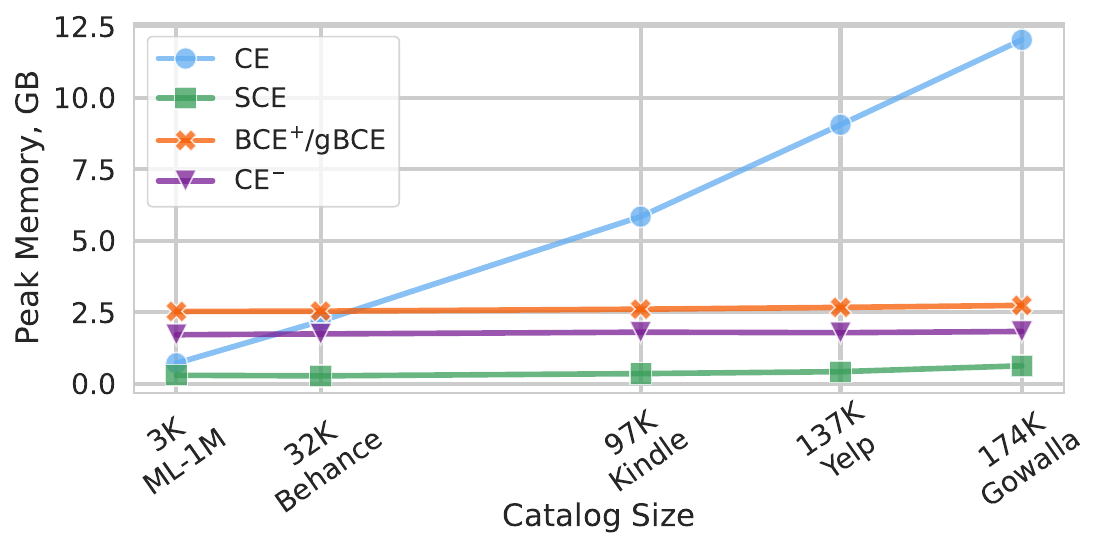}
    \caption{Peak GPU memory utilization during training stage for different catalog sizes. Models trained with batch size equal 64. Models 
 with SCE/BCE$^+$/gBCE/CE$^{-}$ as a loss are trained with 256 negatives.}
    \label{fig:memory_vs_catalog}
    \Description[]{}
\end{figure}

To analyze the effects of catalog size on the memory consumed by various approaches, we trained models with fixed hyperparameters on datasets of different catalog sizes and compiled the results into Figure \ref{fig:memory_vs_catalog}. In this figure, all approaches across the datasets have less than a $10\%$ NDCG@10 difference compared to full CE. In addition to the datasets described in Section \ref{sec:datasets}, we included the widely used benchmark for sequential recommenders \cite{li2021ml, Petrov_2023, wu2021ml}, MovieLens-1M (ML-1M) \cite{harper2015movielens}, which has the smallest catalog among those presented in the figure. Interestingly, we can see that, for the proposed setup, negative sampling-based methods are impractical for datasets with small catalogs, up to $40$K items.  For example, in the case of ML-1M and Behance, with catalog sizes of $3$K and $32$K, respectively, it is more memory-efficient (reducing memory usage by $71.9\%$ and $13.8\%$, respectively) to use full CE instead of BCE$^+$/gBCE. This is because negative sampling methods require storing a significant additional amount of memory for negative sample embeddings.
Furthermore, it is notable that even with modest catalogs, the SCE approach, which employs a Cross-Entropy approximation mechanism distinct from uniform negative sampling, continues to demonstrate superiority over CE in terms of memory usage, with reductions of $-88.4\%$ and $-88.0\%$ compared to BCE$^+$/gBCE for ML-1M and Behance, respectively, which confirms the universality of the proposed method. As the size of the catalog increases beyond $40$K items, the advantage of CE in peak memory consumption relative to negative sampling-based methods begins to diminish. For instance, BCE$^+$/gBCE consumes $77.2\%$ less memory than CE on the Gowalla dataset with $174$K items, and CE$^-$ consumes $84.5\%$ less. This makes the use of BCE$^+$/gBCE/CE$^-$ a viable solution to address memory limitations, on par with the even more memory-efficient SCE ($-93.7\%$). 

\subsubsection{Model Performance Under Memory Constraints}\label{sec:results3}

\begin{figure*}[]
    \centering
    \setlength{\abovecaptionskip}{6pt} 
    \includegraphics[width=1.0\textwidth]{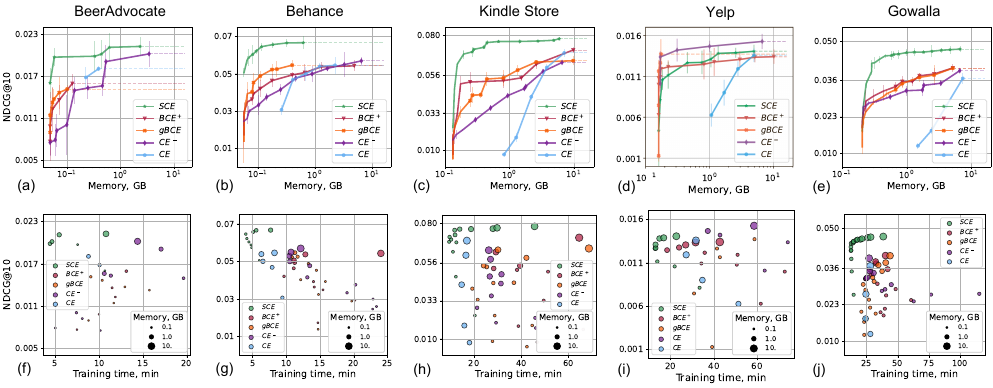}
    \caption{(a)-(e) Pareto front curves with respect to NDCG@10 for a given memory budget. (f)-(j) Same points on the Training time vs. NDCG@10 axes, the size of the point corresponds to the Memory value from the corresponding (a)-(e) plots.}
\label{fig:baselines}
\end{figure*}

To compare with the baselines presented in Section \ref{sec:baselines}, we constructed a grid of values for the hyperparameters that influence peak memory. In addition to batch size, which is a hyperparameter of all methods, for SASRec-BCE$^+$, gSASRec and SASRec-CE$^-$ models, the number of negative samples used for loss calculation acts as such parameter, and for SASRec-SCE it is the number of catalog embeddings in a bucket. We also varied the parameter $t$ of the gSASRec model. The grids for all these hyperparameters can be found in our GitHub repository\footref{github}. Then for each model, we constructed a Pareto front of optimal configurations in terms of achieved NDCG@10 during testing for a given memory budget. Such fronts represent the freedom a model gives to trade memory for quality. These fronts are presented in the first row of Figure \ref{fig:baselines}. Dashed line means that there is no configuration that uses more memory and yields higher NDCG@10. Another important characteristic of a model, besides its memory consumption, is its training time, which reflects both the speed of convergence and the efficiency of the computation.
A better method is located closer to the top left corner of each plot -- this location corresponds to a higher NDCG@10 achieved with less memory and time.

We can see from Figure \ref{fig:baselines} that SASRec-SCE allows to achieve the same quality as the closest competitors with $\textbf{7.8}$ to $\textbf{100.2}$ times less memory on four out of five datasets, while training $\textbf{2.5}$ to $\textbf{6.7}$ times faster. If one is willing to spend more memory, SASRec-SCE gives from $\textbf{3.3}\%$ to $\textbf{17.6}\%$ increase in NDCG@10 compared to the closest competitor on those datasets. The performance on the Yelp dataset is slightly worse and could be potentially explained by the suboptimal selection of heuristic used to choose hyperparameters of our model, described in section \ref{sec:results_hyperparams}, or the fact that Yelp is the most sparse dataset of the datasets we experimented with. We plan to investigate the relationship between the dataset structure and the performance of our method in the future. On Yelp, SASRec-SCE achieves $\textbf{6.2\%}$ worse NDCG@10 than the best-performing model, although training $\textbf{1.9}$ faster than the fastest model yielding the same quality. The extended set of metrics for the best configuration (in terms of NDCG@10) of each model is presented in Table \ref{tab:baselines}.

The reason SASRec-CE model performs poorly on larger datasets despite being a strong baseline is that with memory constraints imposed by the computational device (NVIDIA Tesla V$100$ $16$GB), only a small batch could be processed, which is not enough for good model quality. This aligns with the fact that SASRec-CE is rarely used with large catalogs in practice. 

However, for smaller datasets (BeerAdvocate and Behance), SASRec-CE reaches a plateau in quality before the memory limit is exceeded, so the suboptimal batch size could not explain such underperformance. We see three potential reasons why our method, despite being "just an approximation" of SASRec-CE, still outperforms it even when the latter is not limited by the memory budget. The first possible reason is that in addition to selecting the most informative items in the catalog and thus approximating CE, we also select the most informative states in the sequence and thus deviate significantly from the CE - this feature could help reduce the amount of noise in the input signal. The second potential reason is that the randomized nature of our algorithm helps with generalization due to the general robustness of randomized algorithms \cite{halko2010finding}. Problems with generalization could be significant in the context of recommender systems, given the relatively small sizes of datasets compared to, for example, the sizes of the NLP domain. Finally, it could be possibly connected to the results of the \citet{ma2023exploring}. They suggest that logits with the highest values correspond to so-called False Negatives, i.e., items that should have been marked as positives but were marked as negatives for various reasons, and that sampling these items as negatives has a detrimental influence on model performance. Our model solves MIPS only approximately, so these maximum values could be omitted, and only True Negatives could be found. However, it is difficult to measure the influence of False Negatives on the final metrics and we leave further investigation of sources of improvement of our model for future research.

\begin{table}[]
\setlength{\abovecaptionskip}{6pt} 
\caption{\textbf{{Performance comparison across all datasets. Bold scores are the best on the dataset for the given metric, underlined scores are the second best. Improvements are for SCE compared to the best non-SCE results. NDCG@1 and HR@1 are equivalent by definition.}}} \label{tab:baselines}
\resizebox{\columnwidth}{!}{%
\begin{tabular}{lllllllll}
\hline
Dataset & \multicolumn{2}{l}{Metrics} & \multicolumn{1}{c}{BCE $^+$} & \multicolumn{1}{c}{gBCE} & \multicolumn{1}{c}{CE$^-$} & \multicolumn{1}{c}{CE} & \multicolumn{1}{c}{SCE} & \textit{Improv.} \\ \hline
\multirow{8}{*}{\begin{tabular}[c]{@{}l@{}}Beer\\ Advocate\end{tabular}} & \multirow{3}{*}{NDCG} & @1 & 0.0047 & 0.0045 & \underline{ 0.0082} & 0.0065 & \textbf{0.0097} & \textit{+18.7\%} \\
 &  & @5 & 0.0121 & 0.0113 & \underline{ 0.0163} & 0.0135 & \textbf{0.0174} & \textit{+6.81\%} \\
 &  & @10 & 0.0160 & 0.0152 & \underline{ 0.0202} & 0.0181 & \textbf{0.0209} & \textit{+3.34\%} \\
 \addlinespace[0.6ex]
 & \multirow{2}{*}{HR} & @5 & 0.0201 & 0.0182 & \underline{ 0.0245} & 0.0207 & \textbf{0.0248} & \textit{+1.41\%} \\
 &  & @10 & 0.0322 & 0.0304 & \textbf{0.0366} & 0.0348 & \underline{ 0.0356} & \textit{-2.98\%} \\
  \addlinespace[0.6ex]
 & \multirow{3}{*}{COV} & @1 & 0.0241 & 0.0256 & \underline{ 0.0320} & 0.0305 & \textbf{0.0439} & \textit{+36.9\%} \\
 &  & @5 & 0.0651 & 0.0714 & \underline{ 0.0821} & 0.0795 & \textbf{0.1260} & \textit{+53.7\%} \\
 &  & @10 & 0.0943 & 0.1036 & \underline{ 0.1182} & 0.1149 & \textbf{0.1900} & \textit{+60.1\%} \\ \hline
\multirow{8}{*}{\begin{tabular}[c]{@{}l@{}}Behance\end{tabular}} & \multirow{3}{*}{NDCG} & @1 & 0.0232 & 0.0236 & 0.0243 & \underline{ 0.0248} & \textbf{0.0277} & \textit{+11.6\%} \\
 &  & @5 & 0.0467 & 0.0464 & \underline{ 0.049} & 0.0464 & \textbf{0.0572} & \textit{+16.7\%} \\
 &  & @10 & 0.0543 & 0.0547 & \underline{ 0.0567} & 0.0546 & \textbf{0.0663} & \textit{+16.9\%} \\
 \addlinespace[0.6ex]
 & \multirow{2}{*}{HR} & @5 & 0.0691 & 0.0678 & \underline{ 0.0724} & 0.0672 & \textbf{0.0853} & \textit{+17.9\%} \\
 &  & @10 & 0.0926 & 0.0936 & \underline{ 0.0961} & 0.0927 & \textbf{0.1130} & \textit{+17.9\%} \\
 \addlinespace[0.6ex]
 & \multirow{3}{*}{COV} & @1 & \underline{ 0.0386} & 0.0384 & 0.0368 & 0.0352 & \textbf{0.0393} & \textit{+1.92\%} \\
 &  & @5 & 0.1210 & \underline{ 0.1320} & 0.1280 & 0.1180 & \textbf{0.1530} & \textit{+16.1\%} \\
 &  & @10 & 0.1913 & 0.1960 & \underline{ 0.2002} & 0.1818 & \textbf{0.2500} & \textit{+25.2\%} \\ \hline
\multirow{8}{*}{\begin{tabular}[c]{@{}l@{}}Kindle\\ Store\end{tabular}} & \multirow{3}{*}{NDCG} & @1 & 0.0466 & 0.0440 & 0.0385 & \underline{ 0.0473} & \textbf{0.0560} & \textit{+18.4\%} \\
 &  & @5 & \underline{ 0.0666} & 0.0593 & 0.0581 & 0.0645 & \textbf{0.0737} & \textit{+10.7\%} \\
 &  & @10 & \underline{ 0.0712} & 0.0647 & 0.0638 & 0.0694 & \textbf{0.0782} & \textit{+9.97\%} \\
 \addlinespace[0.6ex]
 & \multirow{2}{*}{HR} & @5 & \underline{ 0.0840} & 0.0732 & 0.0757 & 0.0794 & \textbf{0.0889} & \textit{+5.93\%} \\
 &  & @10 & \underline{ 0.0984} & 0.0898 & 0.0932 & 0.0944 & \textbf{0.1030} & \textit{+4.55\%} \\
 \addlinespace[0.6ex]
 & \multirow{3}{*}{COV} & @1 & \underline{ 0.0379} & 0.0319 & 0.0370 & 0.0336 & \textbf{0.0412} & \textit{+8.82\%} \\
 &  & @5 & \underline{ 0.122} & 0.0952 & 0.1200 & 0.1074 & \textbf{0.1513} & \textit{+23.8\%} \\
 &  & @10 & 0.1830 & 0.1647 & \underline{ 0.1850} & 0.1641 & \textbf{0.2380} & \textit{+28.4\%} \\ \hline
\multirow{8}{*}{Yelp} & \multirow{3}{*}{NDCG} & @1 & 0.0038 & \textbf{0.0043} & \underline{ 0.0042} & 0.0040 & 0.0041 & \textit{-4.22\%} \\
 &  & @5 & 0.0096 & \underline{ 0.0103} & \textbf{0.0107} & 0.0102 & 0.0102 & \textit{-4.54\%} \\
 &  & @10 & 0.0136 & 0.0138 & \textbf{0.0149} & 0.0136 & \underline{ 0.0140} & \textit{-6.21\%} \\
 \addlinespace[0.6ex]
 & \multirow{2}{*}{HR} & @5 & 0.0156 & 0.0163 & \textbf{0.0173} & \underline{ 0.0165} & 0.0165 & \textit{-4.40\%} \\
 &  & @10 & 0.0281 & 0.0273 & \textbf{0.0304} & 0.0272 & \underline{ 0.0281} & \textit{-7.46\%} \\
 \addlinespace[0.6ex]
 & \multirow{3}{*}{COV} & @1 & 0.0119 & 0.0039 & 0.0122 & \underline{ 0.0162} & \textbf{0.0175} & \textit{+7.97\%} \\
 &  & @5 & 0.0364 & 0.0102 & 0.0302 & \underline{ 0.0427} & \textbf{0.0534} & \textit{+25.0\%} \\
 &  & @10 & 0.0590 & 0.0155 & 0.0434 & \underline{ 0.0628} & \textbf{0.0841} & \textit{+33.9\%} \\ \hline
\multirow{8}{*}{Gowalla} & \multirow{3}{*}{NDCG} & @1 & 0.0159 & \underline{ 0.0163} & 0.0158 & 0.0149 & \textbf{0.0207} & \textit{+26.8\%} \\
 &  & @5 & \underline{ 0.0330} & 0.0326 & 0.0321 & 0.0299 & \textbf{0.0393} & \textit{+19.2\%} \\
 &  & @10 & \underline{ 0.0405} & 0.0404 & 0.0396 & 0.0367 & \textbf{0.0476} & \textit{+17.6\%} \\
 \addlinespace[0.6ex]
 & \multirow{2}{*}{HR} & @5 & \underline{ 0.0496} & 0.0486 & 0.0481 & 0.0446 & \textbf{0.0574} & \textit{+15.8\%} \\
 &  & @10 & 0.0730 & \underline{ 0.0731} & 0.0715 & 0.0657 & \textbf{0.0831} & \textit{+13.8\%} \\
 \addlinespace[0.6ex]
 & \multirow{3}{*}{COV} & @1 & 0.0221 & \underline{ 0.0299} & 0.0217 & 0.0235 & \textbf{0.0304} & \textit{+1.77\%} \\
 &  & @5 & 0.0874 & \underline{ 0.1091} & 0.0796 & 0.0767 & \textbf{0.1260} & \textit{+15.2\%} \\
 &  & @10 & 0.1532 & \underline{ 0.1810} & 0.1350 & 0.1231 & \textbf{0.2190} & \textit{+20.8\%} \\ \hline
\end{tabular}%
}
\end{table}

\subsubsection{Evaluating Against Contemporary Models.}\label{sec:results4}

Finally, we compare SASRec-SCE performance on the Amazon Beauty \cite{mcauley2015imagebased} dataset with recently proposed models in the literature, which report the best results. These models include version of SASRec with scalable RECE loss~\cite{gusak2024rece}, three contrastive approaches, CBiT \cite{du2022contrastive}, DuoRec \cite{duo} and CL4SRec \cite{Xie2022ContrastiveLF}, alongside FEARec \cite{fear}, a hybrid attention model that leverages both time-domain and frequency-domain information. 
To make the performance of SASRec-SCE comparable with the performance of these models, we apply corresponding data preprocessing. Furthermore, we diverge from the temporal data-splitting strategy designed to prevent data leakage, described in Section \ref{sec:evals}. Instead, we adopt the widely used leave-one-out approach, where we leave the last interaction for each user in the test dataset, aligning with the evaluation protocols used in the studies of the considered models. Table \ref{tab:recentresults} summarizes this comparative analysis. It demonstrates that SASRec-SCE achieves superior performance among the best recently proposed models.

\begin{table}[]
\setlength{\abovecaptionskip}{3pt} 
\setlength{\belowcaptionskip}{-1pt} 
\caption{\textbf{{Comparison of SASRec-SCE with recent results on Amazon Beauty. Leave-one-out data splitting strategy. Setup follows Petrov et al. \cite{Petrov_2023}. Bold indicates the best value.}}} \label{tab:recentresults}
\resizebox{1.0\columnwidth}{!}{
\begin{tabular}{lcccccc}
\hline
Metric & FEARec & CBiT & DuoRec & CL4SRec & SASRec-RECE & SASRec-SCE \\ \hline
NDCG@10 & 0.0459 & 0.0537 & 0.0443 & 0.0299 & 0.0525 & \textbf{0.0544} \\
HR@10 & 0.0884 & 0.0905 & 0.0845 & 0.0681 & 0.0897& \textbf{0.0935} \\ \hline
\end{tabular}%
}
\end{table}

\section{Conclusion}
\label{sec:conclusion}
\label{sec:conclusion}

In this work, we introduced a novel loss function, Scalable Cross-Entropy, which approximates the Cross-Entropy loss via hard-negative mining using only GPU-efficient operations. With our approach, one can enjoy the benefits of CE loss, which is known to provide state-of-the-art performance, on datasets with large catalog sizes where it is typically not applicable due to its high memory requirements. The scalability improvement of the SCE loss facilitates the training of state-of-the-art Transformer recommender models for potential industrial applications, where datasets comprising millions of items are commonly encountered. We demonstrated that SCE allows us to achieve the same performance as other recently proposed negative sampling methods while using up to $100$ times less memory and training up to $6.7$ times faster on several popular datasets. Alternatively, it could achieve up to $17.6\%$ improvement in quality (NDCG@10) if provided with the same memory budget. In practice, this means that SCE creates a wide range of possibilities for selecting a configuration that meets a user's needs. The underlying principles of SCE can be extended beyond recommender systems models and loss functions, potentially benefiting other domains.\\\\

\bibliographystyle{ACM-Reference-Format}
\balance
\bibliography{recsys_content/7_bib}

\end{document}